\documentclass{iau_FM}
\usepackage{graphicx}

\title[Star formation rate indicators] 
{Panchromatic star formation rate indicators and their uncertainties}

\author[E. da Cunha]   
{Elisabete da Cunha$^1$}

\affiliation{$^1$ Centre for Astrophysics and Supercomputing, Swinburne University of Technology, Hawthorn, VIC 3122, Australia \\email: {\tt edacunha@swin.edu.au}}

\pubyear{2015}
\jname{Astronomy in Focus, Volume 1} 
\editors{Piero Benvenuti, ed.}
\begin{document}

\maketitle

\begin{abstract}
The star formation rate (SFR) is a fundamental property of galaxies and it is crucial to understand the build-up of their stellar content, their chemical evolution, and energetic feedback.
The SFR of galaxies is typically obtained by observing the emission by young stellar populations directly in the ultraviolet, the optical nebular line emission from gas ionized by newly-formed massive stars, the reprocessed emission by dust in the infrared range, or by combining observations at different wavelengths and fitting the full spectral energy distributions of galaxies. In this brief review we describe the assumptions, advantages and limitations of different SFR indicators, and we discuss the most promising SFR indicators for high-redshift studies.
\end{abstract}

Various SFR indicators are commonly used to measure the total rate of ongoing star formation of galaxies from samples at different redshifts (often detected at different wavelengths).
The basic goal is to identify emission from young stars (tracing recent star formation in the last 10 to 100~Myr), and avoid contamination from other sources, such as evolved stellar populations and active galactic nuclei (AGN).
Common assumptions in these calibrations are that the SFR of the galaxy remains constant over a given timescale (often about 100 Myr), that the metallicity is solar, and that the stellar initial mass function (IMF) is know, universal and fully sampled (for extensive reviews, see \cite[Kennicutt~1998, Kenniccutt \& Evans~2012, Calzetti~2013]{Kennicutt1998,Kennicutt2012,Calzetti2012}).

{\em Ultraviolet emission.} The most direct SFR tracer is the ultraviolet (UV) emission emitted by young OB stars, and has the main advantage of being observable in the optical/near-IR bands at high redshift. The calibration of UV SFR indicators depends strongly on the top-end of the IMF, stochasticity of the star formation history (SFH), metallicity, and modelling of the emission by massive stars, which can be complicated by e.g. stellar rotation and binaries. The UV is strongly affected by dust attenuation, and while empirical methods to correct for this have been developed (e.g.~\cite[Meurer et al.~1999]{Meurer1999}), these methods may not be universally applicable (e.g.~\cite[Kong et al.~2004, Dale et al.~2009]{Kong2004,Dale2009}).

{\em Recombination and forbidden lines.} Emission lines from gas ionised by the most massive stars trace current star formation on $<10$~Myr timescales. The intrinsic fluxes of hydrogen recombination lines are known (from case B), making dust corrections easier. However, emission line based SFR indicators suffer from uncertainties related to modelling the most massive stars, stochasticity of SFHs and stochastic IMF sampling at very low SFRs. Additionally, underlying stellar absorption must be accounted for, as well as very high dust optical depths towards the star-forming regions. Finally, there are uncertainties related to possible leakage of ionising photons from HII regions and extra sources of gas excitation (e.g.~shocks, AGN).

{\em Infrared emission by dust.} This traces dust-obscured star formation and is ideal for very dusty sources, or when only IR observations are available. The main uncertainty in calibrating this SFR indicator is the fact that dust can also be heated by old stellar populations (e.g.~\cite[Bendo et al.~2010, Groves et al.~2012]{Bendo2010,Groves2012}). Additionally, we must consider possible dust emission from an AGN torus, variation of IR spectral shapes, and the fact that in many cases not all the emission by young stars is absorbed by dust in galaxies.

{\em Multi-wavelength methods.} In order to circumvent some of the limitations of monochromatic SFR indicators, `hybrid' SFR indicators have been extensively calibrated which combine direct UV (or the H$\alpha$ line) with infrared (or radio) emission (e.g.~\cite[Calzetti et al.~2007, Kennicutt et al.~2009, Hao et al.~2011, Kennicutt \& Evans 2012]{Calzetti2007,Kennicutt2009,Hao2011}), to measure both the unobscured and dust-obscured star formation.
Another way to obtain the SFR of a galaxy is to combine all available multi-wavelength observations through spectral energy distribution (SED) modelling (e.g.~\cite[da Cunha et al.~2008]{daCunha2008}). This method is applicable to a wide variety of galaxies, and has the advantage of breaking degeneracies between dust, age and metallicities, and simultaneously obtaining additional physical parameters such as the stellar mass and dust properties of the galaxies. This method is affected by uncertainties in stellar population modelling, and requires physically realistic descriptions of the SFHs of galaxies (e.g.~\cite[Pacifici et al.~2015]{Pacifici2015}) and of the dust attenuation and emission processes. Given the multi-dimensionality and degeneracies in SED modelling, this method requires sophisticated (e.g.~Bayesian) fitting techniques.

{\em Far-IR fine structure lines.} A promising avenue for high-redshift studies is the calibration of SFR indicators based on fine structure lines such as [CII]158$\mu$m, which present the advantage of being mostly unaffected by dust and easily observable from the ground with modern (sub-)millimetre facilities such as ALMA (e.g.~\cite[Carilli \& Walter 2013]{Carilli2013}). These lines are major coolants of the interstellar medium (ISM), and should correlate with heating, i.e. SFR, however, calibrations are not straightforward and are still ongoing (e.g.~\cite[Ota et al.~2014, Herrera-Camus et al.~2015]{Ota2014,Herrera2015}).

\section*{Acknowledgements}
I would like to thank the organizers of this Focus Meeting for inviting me to this very stimulating meeting and present this review talk, and also
 the International Astronomical Union for financial support to attend the General Assembly.

\end{document}